# Natural criteria of quantum correlations of two coupled oscillators interacting with baths

Illarion Dorofeyev[*]

Institute for Physics of Microstructures, Russian Academy of Sciences,

603950, GSP-105 Nizhny Novgorod, Russia

E-mail: Illarion1955@mail.ru

## Abstract

We consider a problem of description of quantum correlations and dispersions of subsystems of complex open systems. Based on our previous results we proposed a method to evaluate pure quantum contributions from total statistical characteristics of two coupled oscillators interacting with thermal reservoirs. The natural way to extract pure quantum characteristic of the system under study is a calculation of difference between the total value and its pure classical part. A numerical study of temporal behavior of quantum variances and covariances from given initial states up to states in the infinite time limit is given using the proposed method. It is shown that at comparatively high temperatures the steady state total variances are almost classical, because their pure quantum parts tend to zero in this regime. Otherwise, at comparatively low temperatures, the pure quantum parts of variances are not zeroing in steady states manifesting a finite contribution to total values. The same is true for the covariance. The larger the temperature, the lower is the quantum contribution to the total covariance. As well as entanglement measures, quantum discord, the pure quantum contributions in the covariance matrix elements are important characteristics of the temporal dynamics of complex quantum systems.

*) E-mail: Illarion1955@mail.ru

# I. Introduction

The physics of open systems attracts a great interest in different sciences. The key point here is the appropriate model of an open system to describe the most important features of the dynamics of an open system. Very often, the system of selected oscillators coupled to reservoirs of independent oscillators is used as a model of open quantum system [1-4]. Such a model has been successful in describing of various processes, see, for instance [5-9]. In this connection the path integral technique has a special role due to universality [10-13], because the path integration allows obtaining a complete solution to the problem both in quantum and classical limits. The above mention model allows investigating basic phenomena such as an irreversibility in the Brownian dynamics of selected particle within a thermal bath, quantum decoherence of coupled oscillators, temporal behavior of entanglement between two coupled oscillators in contact with baths, temporal dynamics of variances and covariances in different approaches, see introductions in [14-19] for details. In our previous papers, we reported about studies of a bipartite open system constituted by two coupled oscillators interacting with separate reservoirs of harmonic oscillators. We studied a stationary regime of the system in [14], and derived analytical formulas for the mean energy of interaction of the selected oscillators and their mean energies. Temporal dynamics of variances and covariances of two coupled oscillators in separate baths in the weak-coupling limit was investigated in [15]. It was demonstrated that these characteristics in the infinite time limit agree with the Fluctuation dissipation theorem (FDT) despite of initial values. Dynamics of arbitrary coupled identical oscillators was considered in [16]. It was shown that the larger a difference in temperatures of thermal baths, the larger is a difference of the stationary values of variances of coupled identical oscillators as compared to values given by the FDT. The general case of two arbitrary coupled oscillators of arbitrary properties interacting with separate reservoirs is studied in [17]. As well as in previous cases the time-dependent behavior of variances and covariances of oscillators from any initial time up to steady states is investigated. It is shown that the variances and covariances achieve stationary values in the long-time limit. It is demonstrated that the larger the difference in masses and eigenfrequencies of coupled oscillators, the smaller are the deviations of stationary characteristics from those given by the FDT at fixed coupling strength and fixed difference in temperatures between thermal baths. Our paper [18] addresses to study of the relaxation problem of open quantum systems driven by external forces. We considered two bilinear coupled oscillators subjected by arbitrary external forces. In the initial time all interactions among oscillators are switched on and maintained during arbitrary time interval. Then, the arbitrary external forces begin acting at arbitrary time moments. We found and analyzed an analytical expression for time-dependent density matrix for this case. All elements of the covariance matrix are calculated using the known reduction procedure. It is shown that the mean values of coordinates and momenta of coupled oscillators are not zero in case of externally driven oscillators. Time-dependent behavior of the mean values at different conditions is illustrated. Coupled dynamics of selected oscillators at relatively large coupling constants is exemplified at different thermodynamic conditions. It is interesting to note that the mean quadratic characteristics of oscillators are different for the case of "freely developing" pair of oscillators and for the driven pair of oscillators, but their dispersions are identically equal. It should be emphasized that in all above described cases we obtained the density matrix and corresponding covariance matrix of observables, which are valid for quantum and classical regimes.

This paper is devoted to a selection of pure quantum contributions to variances and covariances of two coupled oscillators of arbitrary properties interacting with independent reservoirs. We propose a natural criterion of the quantum contribution as a difference between the total characteristic and its purely classical part, which can be easily obtained from our analytical expressions.

The paper is organized as follows. In Sec.II we state the problem to be solved, present the previous our results being used for the corresponding analysis. A numerical study of a temporal behavior of pure quantum variances and covariances from given initial states up to states in the infinite time limit are given in Sec.III. Final conclusions are given in Sec.IV.

## II. Problem statement.

We consider the system of two coupled oscillators interacting with separate reservoirs of independent harmonic oscillators [14-19]. In case of factorized initial states of two coupled oscillators of arbitrary properties and arbitrary coupling strength, the total density matrix is as follows

$$\begin{aligned}\rho(X_1, X_2, t; \xi_1, \xi_2, 0) &= C(t) \\ &\exp\{-[g_1(t)X_1^2 + g_{12}(t)X_1X_2 + g_2(t)X_2^2]\} \\ &\times \exp\{-[g_1'(t)\xi_1^2 + g_{12}'(t)\xi_1\xi_2 + g_2'(t)\xi_2^2]\} \\ &\times \exp\{i[g_{11}''(t)X_1\xi_1 + g_{21}''(t)X_2\xi_1 + g_{12}''(t)X_1\xi_2 + g_{22}''(t)X_2\xi_2]\}\end{aligned} \qquad (1)$$

where $X_\alpha = x_\alpha + y_\alpha$ and $\xi_\alpha = x_\alpha - y_\alpha$, ($\alpha = 1, 2$), the time-dependent functions $C(t)$, $g(t)$, $g'(t)$ and $g''(t)$ are written in [18,19].

After reduction of the total density matrix with respect to the observables of selected oscillator followed by a normalisation procedure, we obtained total set of elements for the $4 \times 4$ covariance matrix and mean values of observables by means of the usual procedure

$$\begin{aligned}\overline{O}_\alpha &= \int_{-\infty}^{\infty} \lim_{y_\alpha \to x_\alpha} \{\hat{O}_\alpha \rho(x_\alpha, t; y_\alpha, 0)\} dx_\alpha, \quad (\alpha = 1, 2) \\ \overline{O}_{mn} &= \int_{-\infty}^{\infty}\int_{-\infty}^{\infty} \lim_{\substack{y_1 \to x_1 \\ y_2 \to x_2}} \{\hat{O}_{mn} \rho(x_1, x_2, t; y_1, y_2, 0)\} dx_1 dx_2, \quad (m, n = 1, 2)\end{aligned} \qquad (2)$$

where $\hat{O}_\alpha = \{\hat{x}_\alpha, \hat{p}_\alpha, (\hat{x}_\alpha)^2, (\hat{p}_\alpha)^2\}$, ($\alpha = 1, 2$), and

$\hat{O}_{mn} = \{\hat{x}_m \hat{x}_n, \hat{p}_m \hat{p}_n, \hat{x}_m \hat{p}_n, \hat{p}_m \hat{x}_n\}$, ($m, n = 1, 2$), where $\hat{p}_m = -i\hbar \partial / \partial x_m$.

For example, for the dispersion of the coordinate of the first and second oscillators

$$\overline{x_1^2}(t) = \frac{\beta_{22}}{\beta_{11}\beta_{22} - \beta_{12}^2}, \quad \overline{x_2^2}(t) = \frac{\beta_{11}}{\beta_{11}\beta_{22} - \beta_{12}^2}, \qquad (3)$$

where we introduced designations $\beta_{11} = 8g_1(t)$, $\beta_{12} = 4g_{12}(t)$, $\beta_{22} = 8g_2(t)$.

For the coordinate covariance of both oscillators

$$\overline{x_1 x_2}(t) = \frac{\beta_{12}}{\beta_{11}\beta_{22} - \beta_{12}^2}. \qquad (4)$$

It is clear that the total dynamics of coupled oscillators includes both the quantum and classical contributions. The problem is how to extract the pure quantum contributions from the obtained total expressions.

### III. Results and discussion.

#### A. Pure quantum contributions to variances and covariances.

Obviousely, the classical limit of each expression in (3),(4) can be achieved by natural procidure when $\hbar \to 0$. Such a transition yields the single-valued pure classical expressions for variances and covariances. Here, we would like to emphasized advantage of the analytical solution of a problem. Namely, in the obtained expressions for a density matrix in [15-19] it is easy to realize the transition $\hbar \to 0$. Then, we calculate a difference between the total value and corresponding pure classical part. Thus, we derive the seeking for pure quantum contribution of each element of the covariance matrix.

By doing so, we have obtained a total set of 16 matrix elements for the $4\times 4$ covariance matrix. For example, for the pure quantum contributions to variances we have

$$\Delta_i^q(t) = \overline{x_i^2}(t) - \lim_{\hbar \to 0}\left\{\overline{x_i^2}(t)\right\}, \quad (i=1,2), \tag{5}$$

and to pure quantum covariance

$$\Delta_{12}^q(t) = \overline{x_1 x_2}(t) - \lim_{\hbar \to 0}\left\{\overline{x_1 x_2}(t)\right\}. \tag{6}$$

#### B. Quantum dynamics and steady states of spatial variances and covariances of coupled oscillators of arbitrary properties.

For numerical calculations, we have chosen parameters of oscillators as follows $M_1 = 10^{-23} g$ and $\omega_{01} = 10^{13} rad/s$ for the first, and $M_2 = 1.1 M_1$, $\omega_{02} = 1.1\omega_{01}$ for the second oscillator, and $\gamma = 0.01\omega_{01}$ for the dissipative parameter.

The calculations for all figures were done using initial dispersions $\sigma_1^2(t=0) = \hbar/2M_1\omega_{01}$ and $\sigma_2^2(t=0) = \hbar/2M_2\omega_{02}$, which mean the dispersions of isolated oscillators at $T = 0K$.

Figure 1 exemplifies the temporal dependence of the normalized pure quantum parts of variances $\Delta_i^q(t)/\sigma_i^2(FDT)$, $(i=1,2)$ in accordance with Eq. (5) for the first (curve 1) and second (curve 2) oscillators at fixed $\tilde{\lambda} = 0.01$. Here, we consider the system of our study in a state of total thermal equilibrium at $T_1 = T_2 = 300K$ -a) and at $T_1 = T_2 = 30K$ -b). Normalization of $\Delta_i^q(t)$, $(i=1,2)$ was done to equilibrium variances $\sigma_i^2(FDT) \equiv \sigma_i^2(T_i)$, $(i=1,2)$ in accordance with FDT at given temperatures. The normalized coupling parameter is $\tilde{\lambda} = \lambda / \omega_{01}\omega_{02}\sqrt{M_1 M_2}$. It is clearly seen from Fig.1a) that at comparatively high temperatures, when $k_B T_{1,2} \sim \hbar \omega_{01,02}$, the steady state total variances are almost classical, because the pure quantum parts tends to zero at $t \to \infty$. Other behavior is seen at comparatively low temperatures, when $k_B T_{1,2} \ll \hbar \omega_{01,02}$, as it follows from Fig.1b). In this case, the pure quantum parts of variances are not zeroing at $t \to \infty$. Because of distinction in parameters of oscillators, the corresponding distinction in stationary purely quantum contributions to dispersions is noticeable.

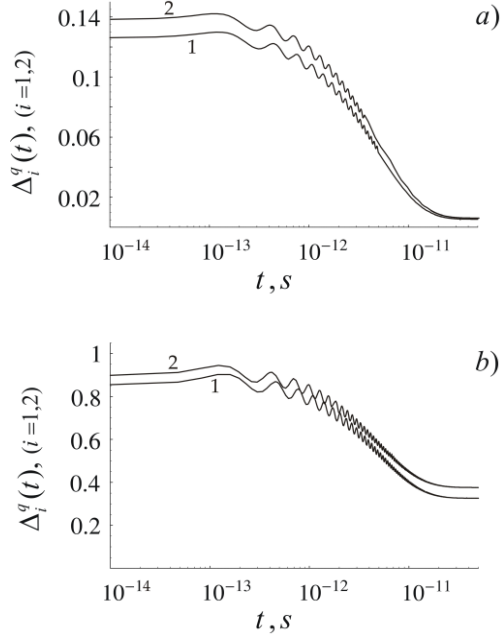
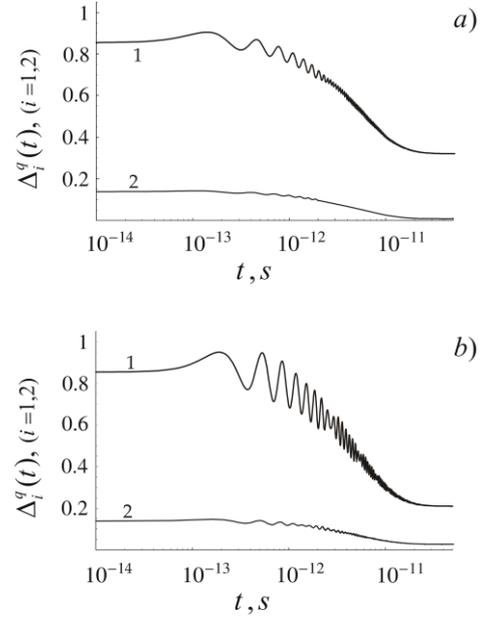

Fig.1  Fig.2

Fig.2 illustrates the temporal dependences of the quantum variances of selected oscillators in a whole system out of total equilibrium in general, when the subsystems in different thermodynamic conditions. In this figure we show the normalized pure quantum parts of variances $\Delta_i^q(t) / \sigma_i^2(FDT)$, $(i=1,2)$ versus a time of the first (curve 1) and second oscillators (curve 2) in accordance with Eq. (5) at $T_1 = 30K$ and $T_2 = 300K$. The variances are calculated at essentially different coupling constants $\tilde{\lambda} = 0.01$ in Fig.2a) and at $\tilde{\lambda} = 0.2$ in Fig.2b). Normalization was done to equilibrium variances in accordance with FDT at $T_1$ and $T_2$ correspondingly. It is seen that the larger the coupling between oscillators the larger is the disturbance of each oscillator during the relaxation process. Besides, the larger the coupling the closer are the steady state values of dispersions due to their mutual influence.

Another very important part of our study connected with quantum effects is a quantum correlation between of subsystems of complex open systems.

Figure 3 shows the temporal dependence of the normalized pure quantum part of covariance $\Delta_{12}^q(t)/\sqrt{\sigma_1^2(FDT)\sigma_2^2(FDT)}$ in accordance with Eq. (6) at $T_1 = T_2 = 300K$ - a) and at $T_1 = T_2 = 30K$ – b). Calculations are done at fixed $\tilde{\lambda} = 0.01$. It is obvious by construction that this value characterizes quantum correlation of coordinates of two oscillators.

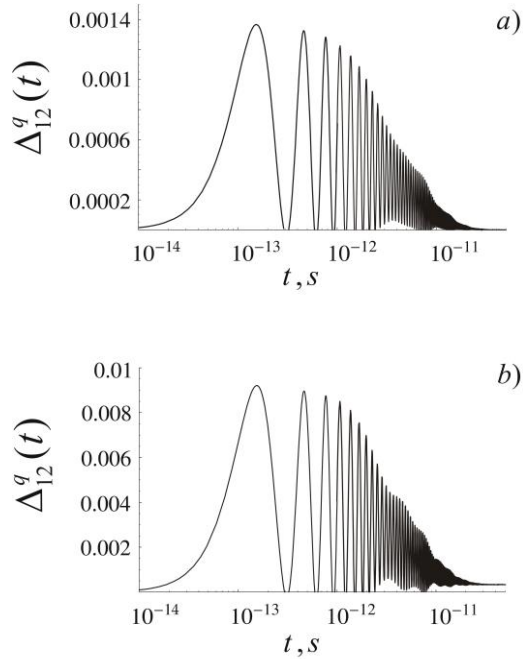

Fig.3

It can be seen from Fig.3a) that at comparatively high temperatures, when $k_B T_{1,2} \sim \hbar \omega_{01,02}$, the quantum correlation of coordinates tends to zero at $t \to \infty$ and the steady state covariance is completely classical. Otherwise, at low temperatures, when $k_B T_{1,2} \ll \hbar \omega_{01,02}$, the quantum correlations are not zeroing at $t \to \infty$ and have finite contributions to the total correlations of coordinates as it follows from Fig.3b).

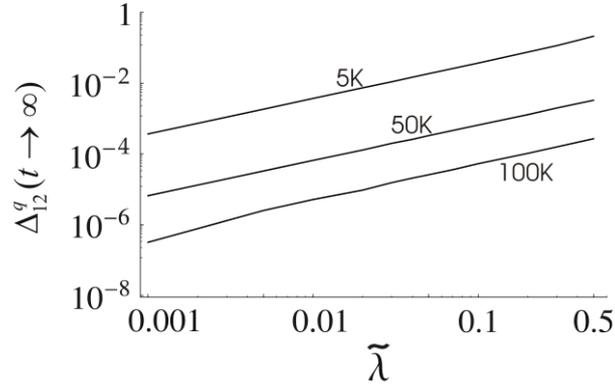

Fig.4

It is also confirmed by the following Fig.4, which shows the stationary normalized pure quantum part of the covariance $\Delta_{12}^q(t=\infty)/\sqrt{\sigma_1^2(FDT)\sigma_2^2(FDT)}$ versus the coupling constant $\tilde{\lambda}$ in accordance with Eq. (6) at different temperatures. The larger the temperature, the lower is the quantum contribution to the total covariance.

## IV. Conclusion.

Our paper addresses the problem of description of quantum correlations and dispersions of subsystems of complex open systems. Based on our previous results we proposed a method to evaluate pure quantum contributions from total statistical characteristics of two coupled oscillators interacting with thermal reservoirs. Then, we studied the relaxation quantum dynamics of the system to quasi-equilibrium states. The natural way to extract the pure quantum characteristic of the system under study is a calculation of difference between the total value and its pure classical part, which can be easily obtained from our analytical expressions in [18, 19]. A numerical study of a temporal behavior of pure quantum variances and covariances from given initial states up to states in the infinite time limit is given. It is shown that at comparatively high temperatures the steady state total variances are almost classical, because the pure quantum parts tend to zero in this regime. Otherwise, at comparatively low temperatures, the pure quantum parts of variances are not zeroing in steady states, manifesting a finite contribution to total values. The same conclusion can be made for the covariance. The larger the temperature, the lower is the quantum contribution to the total covariance. As well as entanglement measures, quantum discord, the pure quantum contributions in the covariance matrix are important characteristics of the temporal dynamics of complex quantum systems. It is necessary to emphasize that the ability to obtain the asymptotic behavior at $\hbar \to 0$ of the complete analytical solution of this problem is obvious advantage in comparison with the numerical solutions.

**Figure captions:**

**Figure 1.** Temporal dependence of the normalized pure quantum parts of variances $\Delta_i^q(t)/\sigma_i^2(FDT)$, $(i=1,2)$ of the first (curve 1) and second (curve 2) oscillators at fixed $\tilde{\lambda}=0.01$ in accordance with Eq. (5) at $T_1=T_2=300K$ - a) and at $T_1=T_2=30K-$ b). Normalization was done to equilibrium variances in accordance with FDT.

**Figure 2.** Temporal dependence of the normalized pure quantum parts of variances $\Delta_i^q(t)/\sigma_i^2(FDT)$, $(i=1,2)$ of the first (curve 1) and second (curve 2) oscillators in accordance with Eq. (5) at $T_1=30K$ and $T_2=300K$ at $\tilde{\lambda}=0.01$ - a) and at $\tilde{\lambda}=0.2$ - b). Normalization was done to equilibrium variances in accordance with FDT at $T_1$ and $T_2$ correspondingly.

**Figure 3.** Temporal dependence of the normalized pure quantum part of covariance $\Delta_{12}^q(t)\big/\sqrt{\sigma_1^2(FDT)\sigma_2^2(FDT)}$ at fixed $\tilde{\lambda}=0.01$ in accordance with Eq. (6) at $T_1=T_2=300K$ - a) and at $T_1=T_2=30K$ – b).

**Figure 4.** Steady state normalized pure quantum part of the covariance $\Delta_{12}^q(t=\infty)\big/\sqrt{\sigma_1^2(FDT)\sigma_2^2(FDT)}$ versus the coupling constant $\tilde{\lambda}$ in accordance with Eq. (6) at $T_1=T_2=5K, 50K, 100K$.